%% file: ms.tex
\documentclass[10pt]{article}

\advance \topmargin by -\headheight
\advance \topmargin by -\headsep
\evensidemargin \oddsidemargin
\usepackage{fullpage}
\usepackage[T1]{fontenc}
\usepackage[utf8x]{inputenc}
\usepackage{palatino}

\newcommand{\cilkpp}{Cilk\texttt{++}}
\newcommand{\ampp}{AM\texttt{++}}
\newcommand{\charmpp}{Charm\texttt{++}}
\newcommand{\cpp}{C\texttt{++}}
\newcommand{\parallex}{ParalleX}
\newcommand{\hpx}{\textsc{HPX-3}}
\newcommand{\xpi}{\textsc{XPI}}

\usepackage{ifthen}

\usepackage{float}
\usepackage{flushend}
\usepackage{color}
\usepackage[usenames,dvipsnames,table]{xcolor}
\definecolor{darkgreen}{rgb}{0.0,0.5,0.0}
\definecolor{darkblue}{rgb}{0.0,0.0,0.5}
\usepackage[pdftex,bookmarks=false,colorlinks=true,linkcolor=darkblue,citecolor=darkblue,urlcolor=darkblue]{hyperref}
\usepackage{url}

\usepackage{graphicx}
\usepackage{soul}
\usepackage{ifpdf}
\ifpdf
  \graphicspath{{./figures/}}
  \DeclareGraphicsExtensions{.pdf}
\else
  \graphicspath{{./figures/}}
  \DeclareGraphicsExtensions{.png,.jpg,.mps,.tif,.eps}
\fi

\usepackage{amsmath}
\usepackage{tabularx}
\usepackage{booktabs}
\usepackage{algorithmic}
\usepackage{array}
\usepackage{mdwmath}
\usepackage{mdwtab}
\usepackage{enumitem}
\usepackage{eqparbox}
\usepackage[subrefformat=parens,labelformat=parens,format=hang,justification=raggedright]{subfig}
\usepackage{fixltx2e}
\usepackage{amstext}
\usepackage{amssymb}
\usepackage{wasysym}
\usepackage{amsfonts}
\usepackage{proof}
\usepackage[ruled,vlined,linesnumbered]{algorithm2e}
\usepackage{listings}
\usepackage{minibox}
\usepackage{multirow}
\usepackage[nameinlink]{cleveref}
\usepackage[draft]{pgf}
\usepackage{textcomp}
\usepackage{fancyvrb}
\usepackage{fancyhdr}
\usepackage{hyperxmp}
\usepackage{rotating}

\hypersetup{colorlinks=true, linkcolor=blue, bookmarksnumbered=true,
  pdfauthor=Abhishek Kulkarni, pdftitle=A Comparative Study of Asynchronous Many-Tasking Runtimes,
  pdfcreator=Abhishek Kulkarni, pdfsubject=Parallel Programming}

\newcolumntype{b}{X}
\newcolumntype{s}{>{\centering\hsize=.25\hsize}X}
\newcommand{\heading}[1]{\multicolumn{1}{|c}{\textbf{#1}}}

\newlength{\circlewidth}
\newcommand{\Topcircle}{\begin{turn}{270}\Leftcircle\end{turn}}
\newcommand{\BOTTOMCIRCLE}{\begin{turn}{270}\RIGHTCIRCLE\end{turn}}
\newcommand{\halfcircle}{\raisebox{1.45ex}{{\Topcircle}\llap{\BOTTOMCIRCLE}}}

\newcommand{\Yes}{\CIRCLE}
\newcommand{\No}{\Circle}
\newcommand{\Maybe}{\halfcircle}

\usepackage[restart]{parnotes}

\begin{document}

\title{A Comparative Study of Asynchronous Many-Tasking Runtimes: Cilk, Charm++, ParalleX and AM++}
\author{Abhishek Kulkarni \quad Andrew Lumsdaine\\
  Center for Research in Extreme Scale Technologies \\
  Indiana University \\
  \href{mailto:adkulkar@crest.iu.edu}{\texttt{\{adkulkar,}}
  \href{mailto:lums@crest.iu.edu}{\texttt{lums\}@crest.iu.edu}}}

\date{January 8, 2014}
\captionsetup{belowskip=2pt,aboveskip=2pt}
\setlength{\intextsep}{2pt plus 2pt}
\maketitle
\lstMakeShortInline[]@

\input{abstract}
\input{body}

\bibliographystyle{abbrv}
\bibliography{ms}
\end{document}

%% file: abstract.tex
\begin{abstract}

We evaluate and compare four contemporary and emerging runtimes for
high-performance computing (HPC) applications: Cilk, \charmpp, \parallex\ and
\ampp. We compare along three bases: programming model, execution model and the
implementation on an underlying machine model. The comparison study includes a
survey of each runtime system's programming models, their corresponding
execution models, their stated features, and performance and productivity
goals.
    
We first qualitatively compare these runtimes with programmability in mind. The
differences in expressivity and programmability arising from their syntax and
semantics are clearly enunciated through examples common to all runtimes. Then,
the execution model of each runtime, which acts as a bridge between the
programming model and the underlying machine model, is compared and contrasted
to that of the others. We also evaluate four mature implementations of these
runtimes, namely: Intel \cilkpp, \charmpp\ 6.5.1, \ampp\ and \hpx, that embody
the principles dictated by these models.
  
With the emergence of the next generation of supercomputers, it is imperative
for parallel programming models to evolve and address the integral challenges
introduced by the increasing scale. Rather than picking a \textit{winner} out
of these four models under consideration, we end with a discussion on
\textit{lessons learned}, and how such a study is instructive in the evolution
of parallel programming frameworks to address the said challenges.

\end{abstract}


%% file: body.tex

\section{Introduction}
\label{sec:intro}

In June 2013, the Tianhe-2, a supercomputer developed by China's National
University of Defense Technology, attained a performance of 33.86 Petaflop/s on
the LINPACK benchmark. At the time, there were at least 26 systems that have
reached the Petaflop mark ($10^{15}$ floating-point operations per
second)~\cite{Top500}. It has been predicted that by the end of this decade,
supercomputers will reach the \textit{exascale} mark with potentially thousands
of nodes and hardware accelerators, millions of cores, and billions of threads
of execution~\cite{Exascale:DARPA}. This pinnacle of high-performance computing
capability would, unarguably, help advance scientific discoveries and benefit
diverse domains such as climate and ocean modeling, national security, energy,
materials science, etc.

There are several key challenges in how we build computing systems, and how we
develop system software that need to be addressed before supercomputers can
attain sustained \textit{exascale} performance. The four most significant
challenges towards reaching this target are: Energy and Power; Memory and
Storage; Concurrency and Locality; and Resiliency~\cite{Exascale:DARPA}.

The scalability of applications that are expected to benefit from the computing
capability offered by these exascale-class machines largely depends, among
other things, on programmability of these machines. Programmability is
primarily dictated by the parallel programming model exposed by the de-facto
runtime on these machines to the programmer. Several parallel programming
frameworks exist, but are encumbered by their limitations in expressing
parallelism, or by the inadequacy of their corresponding runtime systems to
maintain efficiency at higher scales.



A parallel programming framework typically consists of a programming
model (the language in which a parallel program is expressed), a
corresponding execution model (that defines \emph{how} that program is
executed) with the help of an underlying implementation based on a
given machine model. The language and its associated runtime comprise
a parallel programming framework. Several parallel programming
frameworks are in prevalent use today, including shared-memory
programming models such as Pthreads, OpenMP~\cite{OpenMPSpec},
Cilk~\cite{cilk-1996,Frigo:1998:ICM:277650.277725}, Intel (Threading
Building Blocks) TBB~\cite{tbb,tbb-reference-manual};
accelerator-based programming models such as CUDA~\cite{cuda},
OpenCL~\cite{opencl,opencl-spec} and OpenACC; and distributed-memory
programming models such as the Message Passing Interface
(MPI)~\cite{MPI-2.2,MPI-3.0}, MapReduce~\cite{MapReduce},
\charmpp~\cite{CharmppOOPSLA93}, \parallex~\cite{Parallex},
UPC~\cite{upc}, X10~\cite{Charles:2005:XOA:1094811.1094852} and
others. Many of these frameworks have experienced a wide adoption
(some more than others) in the programming community. In addition to
allowing ease of programmability, another prime concern of parallel
programming frameworks is increased efficiency and higher
performance. The higher performance is attained by exploiting
opportunities offered by the advances in hardware architecture.

In this report, we compare four distinct, but related, parallel
programming frameworks for high-performance computing: Cilk,
\charmpp, \ampp\ and \parallex. Our comparison spans three bases:
programming model, execution model, and their realizations: an
underlying implementation. We evaluate along these axes, since the
original goals of these frameworks is both, high productivity and high
performance. If we were only interested in \textit{one} of the two,
simply looking at the programming model (syntax) or the performance of
the implementation would have sufficed. Evaluating along these
dimensions provides us with a fair \textit{apples-to-apples}
comparison of Cilk, \charmpp, \ampp\ and \parallex.

\begin{figure}[!ht]
\centering
\includegraphics[scale=0.6]{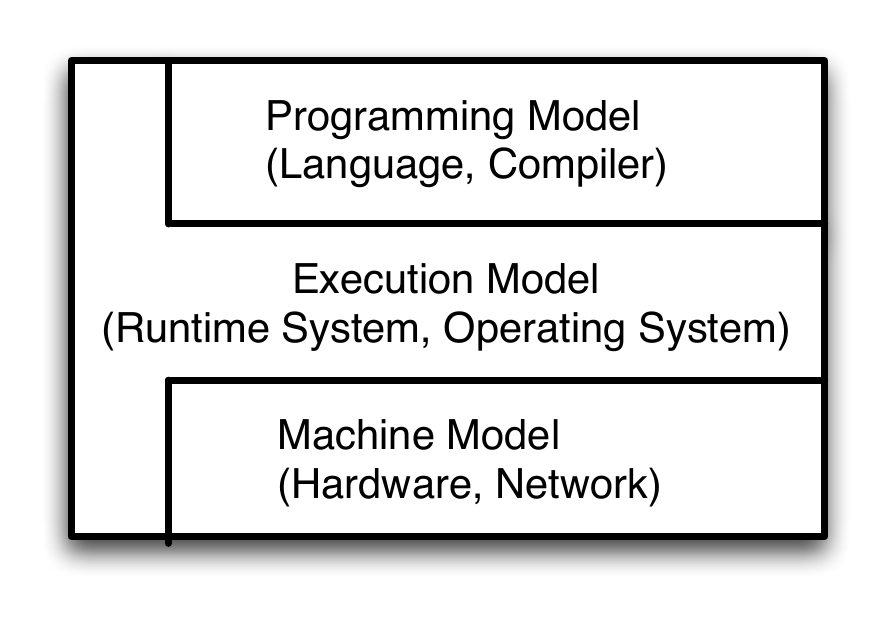}
\caption{Components of a Parallel Programming Framework.}
\label{fig:parprog}
\end{figure}

\paragraph{Programming Model}

A parallel programming model provides the constructs for exposing and
expressing latent parallelism in a program. Existing parallel
programming paradigms are mainly characterized based on the inherent
data- or control-parallelism that they expose. Traditional parallel
programming paradigms widely in adoption are: SIMD/SPMD
(single-program multiple-data), which helps expose data-parallelism;
MIMD/MPMD (multiple-program multiple-data), which helps expose
task-parallelism; and asynchronous execution models such as dataflow,
and message-driven execution, which expose both data- and
control-parallelism.

\begin{figure}[!ht]
\centering
\noindent\begin{minipage}{.42\textwidth}
\begin{lstlisting}[caption=OpenMP]{OpenMP}
const int N = 100000;
int i, a[N], b[N], c[N];
 
#pragma omp parallel for
for (i=0; i<N; i++)
  a[i] = b[i] = 2*i;

#pragma omp parallel for \
 shared(a,b,c,100) private(i) \
 schedule(static,100)
for (i=0; i<N; i++)
  c[i] = a[i] + b[i];
\end{lstlisting}
\end{minipage}\hfill
\noindent\begin{minipage}{.55\textwidth}
\begin{lstlisting}[caption=Pthreads]{Pthreads}
const int N = 100000;
int i, a[N], b[N], c[N];
int args[NUM_CPUS];
pthread_t threads[NUM_CPUS];

void *init(void *arg) {
  int tid = *((int*) arg);
  int range = N/NUM_CPUS;
  for (i=0; i<range; i++)
    a[i] = b[i] = (range*tid)+2*i;
}

// void *sum(void *arg) { ... }

for (i=0; i<NUM_CPUS; ++i) {
  args[i] = i;
  pthread_create(&threads[i],0, init,
                 (void*)&args[i]);
}

for (i=0; i<NUM_CPUS; ++i)
  pthread_join(threads[i],0);

for (i=0; i<NUM_CPUS; ++i)
  pthread_create(&threads[i],0, sum,
                 (void*)&args[i]);

for (i=0; i<NUM_CPUS; ++i)
  pthread_join(threads[i],0);
\end{lstlisting}
\end{minipage}
\caption{Data-parallel vector-addition expressed in OpenMP and POSIX Threads.}
\label{fig:openmp-pthreads}
\end{figure}

From a programmer's perspective, the other differentiating factor of
programming models is based on how they abstract the programmer's view
of memory (and, thus, data). Contemporary parallel programming
frameworks can be predominantly divided into two major classes:
message-passing models based on the CSP (Communicating Sequential
Processes) execution model, and shared-memory models which provide a
uniform, global, shared-view of the memory.

A \emph{programming model} enables the expression of parallelism in a
given program. Productivity of a programming model is, thus, dependent
on the overall expressivity of its language. Higher-level programming
models leave out a lot of mundane details, that are filled in by the
runtime, in favor of higher expressivity (e.g. OpenMP). On the other
hand, low-level programming models such as POSIX threads provide full
control to the programmer thus allowing to get the maximum performance
out of them. Oftentimes, high-level programming paradigms use
low-level programming paradigms as their targets to offer the best of
both worlds. \Cref{fig:openmp-pthreads} shows a data-parallel
vector-addition example expressed in a high-level programming framework
(OpenMP) in contrast to a low-level programming abstraction
(Pthreads). The details about task granularity, data locality etc. are
either left out or can be specified through domain-specific
annotations.

The fundamental characteristics of a parallel programming model are:
\begin{enumerate}[leftmargin=*]
  \item Linguistic constructs for parallel control (@spawn@, @doall@).
  \item Programmer's view of memory (shared memory, message passing).
  \item Explicit control over data layout or distribution.
  \item Abstraction and encapsulation of control flow (synchronous, asynchronous).
  \item Reuse, modularity and composability.
  \item Determinism and repeatability v/s non-determinism.
\end{enumerate}

Since a programming model offers a language (or language extension
through annotations) to express parallelism, it often requires a
compiler (or a preprocessor or source-to-source translator) to
translate the language into instructions that the execution model can
execute. Although, this requires additional development effort, it
admits domain-specific compile-time optimizations and static analyses
that can guide the runtime system to achieve higher performance. Such
domain-specific languages have shown to improve programmer
productivity, while maintaining all of the performance benefits.



\paragraph{Execution Model}
As the programming model is said to introduce syntax for parallelism
in a programming language, the corresponding execution model
complements it by providing an operational semantics for the
syntax. Put simply, while the programming model enables \emph{what} is
to be expressed, the execution model dictates \emph{how} it would be
executed. An execution model is thus an abstraction of computation
that also captures the inter-relationships between programming model
and the underlying machine model. A parallel execution model takes
into account communication, locality, synchronization etc. to turn
latent parallelism into actual parallelism with maximal efficiency.

The defining characteristics of a distributed execution model are:
\begin{enumerate}[leftmargin=*]
  \item Units of computation (threads, tasks).
  \item Communication of data between tasks (single-node v/s
    multi-node).
  \item Synchronization between tasks (through communication).
  \item Abstraction and encapsulation of control flow (synchronous v/s asynchronous).
  \item Efficient execution of tasks on a machine (scheduling, mapping etc.).
  \item Runtime features such as termination detection, quiescence
    detection, parallel I/O, fault tolerance, checkpoint/restart etc.
\end{enumerate}

An execution model conceptually spans the entire vertical stack
defining, largely, the features exposed by the programming model to
the programmer and relying heavily on the features exposed by the
machine to the runtime and operating systems. A programming model can,
sometimes, be quite similar to the underlying execution model (with a
superficial difference between the two, at times). 

Similarly, some execution models are a better (and more natural) fit
for some machine models: e.g. data-parallel models on vector
processors, message-driven execution models on dataflow machines
etc. Execution models such as MPI, which is based on the Communicating
Sequential Processes (CSP) model have enjoyed a much broader adoption,
whereas upcoming execution models such as those offered by the
Partitioned Global Address Space (PGAS) languages like UPC and
Co-Array Fortran have been catching up lately.

\paragraph{Implementation}
The execution model, typically, depends on an abstract machine model by
parameterizing over the nitty-gritty machine details. Different
underlying architectures, however, enable optimizations that can
improve the overall performance of the parallel programming
model. Existing supercomputers comprise of a wide variety of machine
architectures (some homogeneous, and others heterogeneous), and
varying network characteristics. An implementation of a parallel
programming framework is, thus, characterized by factors such as

\begin{itemize}[leftmargin=*]
  \item Efficiency of the implementation (Performance).
  \item Portability (Different network transports).
  \item Interoperability (compatibility with other runtime systems).
  \item Tools and infrastructure (debugging, visualization).
\end{itemize}

Oftentimes, the implementations offer valuable insight into the
shortcomings of an execution model, and help in extending the
programming framework to plug those. For instance, the Intel Cilk Plus
extensions define enhancements to the original Cilk model (reducers,
array notation etc.) that capture some of the common information
sharing patterns for task-parallelism.

%

Our main contributions in this paper are as follows:

\begin{itemize}[leftmargin=*]
\item We identify the issues concerning extreme-scale computing and
  how execution models address the challenges posed by some of these
  issues. We categorize a parallel programming framework into a
  programming model, an execution model, and an implementation.
\item We qualitatively compare three parallel programming frameworks
  with respect to the proposed categorization.
\item We provide a discussion to evolve and extend the \parallex\
  execution model to address some of its limitations.
\end{itemize}

The remainder of this paper is structured as follows: \Cref{sec:cilk}
describes the Cilk programming framework through a variety of
examples. \Cref{sec:charm} presents \charmpp, a message-driven,
object-oriented, asynchronous execution model for high-performance
computing. \Cref{sec:parallex} introduces \parallex, a
high-performance parallel execution model that is built upon
fundamental principles that could potentially subsume multiple
execution paradigms to achieve scalability. \Cref{sec:comparison}
compares three featureful implementations of these execution
models. Finally, in \Cref{sec:discussion} we conclude with a
discussion and mention the future directions for this study.


\section{Cilk}
\label{sec:cilk}

Cilk is a multi-threaded language for parallel programming that
introduces syntax for expressing parallel
control~\cite{Frigo:1998:ICM:277650.277725}. It is provided as an
extension to the C (and \cpp) language to support data and task
parallelism. The original Cilk model, at MIT, was proposed and
implemented as early as 1994. Since then Cilk has undergone several
revisions, and the most commonly used Cilk implementation today, Intel
Cilk Plus, is based on the original Cilk 5 model. The Cilk philosophy
was to provide a \emph{parallel} extension to C, and hence, it ensures
sequential semantics when the Cilk keywords are ignored. This property
is called as ``C elision'' (or more generally ``serial
elision''). Further, Cilk implements an efficient work-stealing
scheduler providing nearly optimal scheduling of parallel tasks. Intel
has further introduced extensions to the orignal Cilk model with two
key constructs: \emph{reducers} - which are shared objects that allow
two tasks to communicate without introducing a data-race, and
\emph{array notation} - which allows leveraging vector capabilities of
modern processors.

\subsection{Programming Model}

Cilk introduces two keywords for expressing \textit{logical
  parallelism} within a program: @cilk_spawn@ and
@cilk_for@. Synchronization of tasks is expressed through the keyword
@cilk_sync@. Originally Cilk also allowed specification of
non-determinism through the keywords @abort@ and @inlet@.

The spawn keyword slight differs from a traditional thread spawn, in
that, it only expresses the intent of parallelism to the Cilk
runtime. It differs from a traditional C function call as the function
annotated with the @cilk_spawn@ keyword can now be executed in
parallel with the parent function. Thus, it represents an opportunity
for parallelism, and is not a command that enforces parallelism. It is
not safe to use the values returned by the children until they execute
a @cilk_sync@ statement. @cilk_spawn@ specifies that all spawned calls
in a function complete before the execution continues. It can be seen
as a ``local barrier'' for the given function. There is an implicit
@cilk_sync@ at the end of every function and every try block that
contains a @cilk_spawn@. When the spawned function returns, the values
are stored directly in the parent's frame. However, previous Cilk
versions provided an @inlet@ call, which allowed the incorporation of
the returned value in to the parent's frame in a more complicated
way. The @inlet@ can be viewed as a one-shot continuation that is
executed after the child returns. The example in \Cref{fig:cilk-fib}
shows a parallel version written in different versions of Cilk.

\begin{figure}[!ht]
\centering
\noindent\begin{minipage}{.5\textwidth}
\begin{lstlisting}[caption=Cilk-2,label=lst:cilk2]
thread Fib(cont int k, int n) {
  if (n < 2)
    send_argument(k, n);
  else {
  cont int x, y;
  spawn_next Sum(k, ?x, ?y);
  spawn fib(x, n-1);
  spawn fib(x, n-2);
  }
}
thread Sum(cont int k,int x,int y) {
  send_argument(k, x+y);
}
\end{lstlisting}\vspace{-5mm}
\begin{lstlisting}[caption=Cilk-5,label=lst:cilk5]
int fib(int n) {
  if (n < 2)
    return n;
  int x = cilk_spawn fib(n-1);
  int y = fib(n-2);
  cilk_sync;
  return x + y;
}
\end{lstlisting}
\end{minipage}\hfill
\noindent\begin{minipage}{.4\textwidth}
\begin{lstlisting}[caption=Cilk with inlets,label=lst:cilk-inlet]
int fib(int n) {
  int x = 0;
  inlet void sum(int s) {
    x += s;
    return;
  }

  if (n < 2)
    return n;
  else {
    sum(cilk_spawn fib(n-1));
    sum(cilk_spawn fib(n-2));
    cilk_sync;
    return x;
  }
}
\end{lstlisting}
\end{minipage}
\caption{A Cilk program to compute the \textit{N}th Fibonacci number.}
\label{fig:cilk-fib}
\end{figure}

The performance of a Cilk computation can be characterized by:
\textbf{work}, which is the serial execution time, and
\textbf{critical-path length}, which is the execution time on an
infinite number of processors. Much of the design influences for Cilk
5 were dictated by the following principle:

\begin{quote}
  \textbf{The work-first principle:} Minimize the scheduling overhead
  borne by the work of a computation. Specifically, move overheads out
  of the work and onto the critical path.
\end{quote}

The total execution time of a Cilk computation on $P$ processors is
bounded by
$$
T_P \leq \frac{T_1}{P} + c_{\infty} T_{\infty}
$$
where the first term represents the \textit{work overhead} and the
second term denotes the \textit{critical-path overhead}. The
work-first principle aims to minimize $T_1$ at the expense of a larger
$c_{\infty}$ as it has a more direct impact on the performance.

Since Cilk relies on a provably-optimal work-stealing scheduler, the
programmer does not have to worry about scheduling or explicit
relocation of threads. In the orignal Cilk model, tasks had to use
explicit locking to synchronize access to shared data. Newer
extensions to Cilk allow race-free access to shared data. Prior Cilk
models also permitted distributed-memory programming allowing a shared
view to global data. Globally shared data was annotated with the
@shared@ keywords, as shown in \Cref{fig:cilk-mmult}. The semantics of
this underlying distributed-shared
memory~\cite{Blumofe:1996:DDS:645606.661333} is discussed in the next
section.

\begin{figure}[!ht]
\centering
\begin{lstlisting}
cilk void matrixmul(long nb, shared block *A,
                             shared block *B,
                             shared block *R) {
  if (nb == 1)
    multiply_block(A, B, R);
  else {
    shared block *C,*D,*E,*F,*G,*H,*I,*J;
    shared block *CG,*CH,*EG,*EH,
                 *DI,*DJ,*FI,*FJ;
    shared page_aligned block tmp[nb*nb];
    /* get pointers to input submatrices */
    partition(nb, A, &C, &D, &E, &F);
    partition(nb, B, &G, &H, &I, &J);
    /* get pointers to result submatrices */
    partition(nb, R, &CG, &CH, &EG, &EH);
    partition(nb, tmp, &DI, &DJ, &FI, &FJ);
    /* solve subproblems recursively */
    spawn matrixmul(nb/2, C, G, CG);
    spawn matrixmul(nb/2, C, H, CH);
    spawn matrixmul(nb/2, E, H, EH);
    spawn matrixmul(nb/2, E, G, EG);
    spawn matrixmul(nb/2, D, I, DI);
    spawn matrixmul(nb/2, D, J, DJ);
    spawn matrixmul(nb/2, F, J, FJ);
    spawn matrixmul(nb/2, F, I, FI);
    sync;
    /* add results together into R */
    spawn matrixadd(nb, tmp, R);
    sync;
  }
  return;
}
\end{lstlisting}
\caption{Matrix Multiplication in Cilk using distributed-shared memory.}
\label{fig:cilk-mmult}
\end{figure}


\subsection{Execution Model}

In Cilk, tasks represent \textit{logical parallelism} in the
program. Since Cilk is a \emph{faithful} extension of C (and \cpp),
the C elision of a Cilk program is a correct implementation of the
sequential semantics of the program. Besides, Cilk exclusively uses
work-stealing as the scheduling strategy in accordance with the
work-first principle. A Cilk DAG represents the series-parallel
structure of the execution of a Cilk program. Nodes of this task DAG
are stolen dynamically without any a priori partitioning. Due to these
reasons, the Cilk runtime creates two version of each Cilk function: a
sequential ``slow'' clone, and a parallel ``fast'' clone. The ``fast''
clones are stolen by idle threads on different processors and executed
as a lightweight-thread with a stack. Due to the C elision of a Cilk
program, the stolen children are backed by ``cactus'' stacks where the
parent stack is common to all of the children spawned by a given
function. The Cilk execution model always executes a spawned function
on the same worker (and presumably the same system thread) whereas the
continuation is stolen by a different worker (and executed,
presumably, by a different thread on a different processor).

The corner stone of the Cilk parallel programming model is its
provably-optimal scheduling based on work-stealing. This scheduling
strategy is based on a prior theoretical analysis of Cilk's execution
model. The runtime load-balancing scheduler implements a
Dijkstra-like, shared-memory, mutual-exclusion protocol (THE protocol)
guaranteeing that \textit{stealing} only contributes to the
critical-path overhead. This protocol mandates that thieves steal from
the head of the queue (contributing to the critical-path overhead)
whereas workers steal from the tail of the shared-memory task queue
(adding only to the work overhead). The workers resort to heavy-weight
hardware locks only when a conflict is detected.

The Cilk multithreaded runtime system originally developed for the
Connection Machine CM5 had support for distributed shared memory
implemented in software. It had a weaker consistency model called as
DAG-Consistent distributed-shared memory. DAG consistency allowed
different reads to return values based on different serial orders, but
the reads respected the dependencies in the DAG. Thus, a read can
``see'' the write only if there is some serial execution order in the
DAG where the ``read'' sees the ``write''. The coherence was
maintained by the \textsf{BACKER} coherence algorithm and is described
in ~\cite{Blumofe:1996:DDS:645606.661333}. Further extensions to this
model involved implementing Lazy Release Consistency (LRC), as in
TreadMarks~\cite{Keleher94treadmarks:distributed}, to improve the
efficiency of the memory consistency model. This system,
SilkRoad~\cite{Peng00silkroad:a}, delayed the propagation of
modifications until the next lock acquisition, thus reducing the
communication cost. Cilk-NOW~\cite{Blumofe:1997:ARP:1268680.1268690}
was an extension to the Cilk-2 language which was largely functional
and used explicit continuations to represent parallelism. More
specifically, the Cilk-2 runtime system did not support putting values
directly into the parent's frame, and hence the parallel program had
to be expressed in a continuation-passing style. Tasks were
represented as heap-allocated closures. An example of a Fibonacci
program written in the Cilk-2 language is shown in
Listing~\ref{lst:cilk2} in \Cref{fig:cilk-fib}. The Cilk-NOW system
extend the Cilk execution model with two features 1) adaptivity - new
nodes could be added (or removed) dynamically to (or from) the
distributed runtime system and 2) reliability - failing nodes did not
affect the running distributed Cilk computation.

\subsection{Implementation}

Intel Cilk Plus is the most widely used existing Cilk
implementation. It is commercially implemented as part of the Intel
C++ Composer XE compiler. In addition to that, open-source
implementations for GCC and LLVM are available from
\url{http://cilkplus.org/download}.


The @cilk_for@ keyword converts a for loop into a parallel for
loop. Each of the loop iteration can be executed in parallel . The
programmer can control the granularity by setting the grain size of
the loop (similar to OpenMP parallel for).

To avoid using locks to synchronize access to shared data, Intel Cilk
Plus offers \textit{reducers} that allows tasks to use private "views"
of a variable which are merged at the next sync. An ordered merge
(over the reducer monoid) ensures the serial semantics of the Cilk
program.

The array notation allow users to express high-level vector operations
on entire arrays or their slices. @arr[:]@ represents an entire array
of stride 1. Similarly, multi-dimensional arrays can be referred to by
@arr[:][:]@. The array notation can be used with both, statically and
dynamically allocated arrays. Several functors are provided that
natively operate on these arrays. It can also be used to represent
scatter and gather operations, like @C[:] = A[B[:]]@.

An elemental function is a function which can be invoked either on
scalar arguments or on array elements in parallel. The Cilk compiler
generates both scalar and vector versions of these functions. The
@#pragma simd@ keyword forces the compiler to generate vectorized code
for a given function.


\section{\charmpp}
\label{sec:charm}

\charmpp\ is a parallel programming system based on message-driven
execution of migratable objects~\cite{CharmppOOPSLA93}. While the
original Charm execution model was implemented as an extension to C,
current \charmpp\ extensions are additions to the \cpp\ language with
a distributed, adaptive runtime system. \charmpp\ can express both
data and task parallelism in a single application. Task parallelism is
primarily expressed through migratable objects (called as
\textit{chares}) whereas data parallelism is exposed through a
collection of these chare elements called as ``chare arrays''. The
\charmpp\ programming model has been demonstrated to perform better
for irregular and latency-sensitive applications, that benefit from
its message-driven execution.

\subsection{Programming Model}

The most important underlying idea behind the \charmpp\ programming
model comes from the actor programming models and dataflow execution
models. The CSP execution model (and its primary realization, MPI)
impose a Bulk Synchronous Parallel (BSP) programming regime which has
shown to incur considerable overheads at higher scales due to strong
synchronization. \charmpp\ like Cilk relies on the parallel slackness
property: that the average parallelism inherent in the application
exceeds much more than the number of processors that it is executed
on. This over-decomposition allows the \charmpp\ runtime system to
employ intelligent load-balancing schemes to migrate objects in the
system~\cite{CharmSys1TPDS94}.

\textbf{Chares} are essentially concurrent objects with methods that
can be invoked remotely. A program is divided into logical tasks
represented by chare classes which are then invoked through message
passing. A \charmpp\ program supports supports modularity and parallel
composition. Modules can depend on other external modules, and also
include additional header files. The interfaces are written in a
``.ci'' file which is preprocessed to generate a \cpp\ header file and
linked with the corresponding \cpp\ code. The extended language
supports multiple inheritance, dynamic binding, overloading and strong
typing.

\paragraph{Chare Definitions}
Chares disallow unrestricted global variables and static variables in
classes. Chare classes have one or more \textbf{entry methods} that
take a list of parameters or message types. Some example declarations
of entry methods are shown below.

\begin{lstlisting}
entry void entryA(int n, float arr[n]);
entry void entryB(int n, float arr[n*n]);
entry void entryC(int nd,int dims[nd],float arr[product(dims,nd)]);
\end{lstlisting}

Note that the entry method arguments can include array references that
are passed by value. Arbitrary message types or user-defined data
structures are also supported as long as they provide the
corresponding serialization and deserialization routines.

All chare objects are mandated to have a constructor entry method, and
any number of other entry methods. Entry methods of a chare are
non-preemptible. A chare's entry methods can be invoked via
proxies. The @thisProxy@ member variable returns the proxy pointer to
itself which can then be passed to other chare objects.

\begin{lstlisting}
// interface.ci
chare ChareType
{
  entry ChareType(parameters1);
  entry void EntryMethodName(parameters2);
};

// interface.h
class ChareType : public CBase_ChareType {
  // Data and member functions as in C++
  public:
    ChareType(parameters1);
    void EntryMethodName2(parameters2);
};
\end{lstlisting}

\paragraph{Chare Creation}
New chares are created through the @ckNew@ method as shown below:
\begin{lstlisting}
CProxy_chareType::ckNew(parameters, int destPE);  
\end{lstlisting}

The @destPE@ parameter is optional, as the runtime migrates the object
to a processing element (PE) based on the load-balancing policy in
effect. Note that chare creation is lazy and asynchronous. 

\paragraph{Method Invocation}

The following call invokes a method @EntryMethod@ on the chare
represented by the proxy @chareProxy@. It is also possible to check if
a proxy points to a local chare or not. This is typically used as an
optimization where local methods can be invoked directly without
forwarding them through a proxy.

\begin{lstlisting}
chareProxy.EntryMethod(parameters)
// check if local
C *c=chareProxy.ckLocal();
\end{lstlisting}

\paragraph{Data objects}

The original Charm model supported four different kind of data
objects, namely: 

\begin{itemize}
\item Read-Only objects.
\item Write-Once objects.
\item Accumulator objects.
\item Monotonic objects.
\end{itemize}

Earlier \charmpp\ versions also supported a Multi-phase Shared Arrays
(MSA)~\cite{desouza04:_msa,MSA-APGAS09} abstraction that layered PGAS
semantics on top of the asynchronous \charmpp\ runtime.

\charmpp\ only supports read-only variables; however, the other
modalities are subsumed by allowing arbitrary reductions on
\textit{chare arrays}. As the name suggests, it is erroneous to assign
to a read-only variable more than once. These are declared using the
@readonly@ annotation as follows @readonly Type VarName;@.

\textbf{Chare arrays} are arbitrarily-sized collections of chares that
are distributed across different procesing elements. They have a
globally unique identifier of type @CkArrayID@, and each element has a
unique index of type @CkArrayIndex@. Chare arrays are declared using
the following syntax:

\begin{lstlisting}
  array [1D] Arr {
    entry Arr(init);
    entry void Size();
  };
  class A : public CBase_Arr {
    public:
      Arr(init);
      Arr(CkMigrateMessage *); // migration constructor
      void someEntry();
  };      
\end{lstlisting}

The @thisProxy@ member variable can be used to return a proxy pointer
to the entire chare array.

\paragraph{Operations on Chare Arrays}
Since chare-arrays are a collection of chare objects, aggregate
operations representing common communication patterns can be performed
on these collections. Method invocation on a particular chare in a
chare-array is simply performed by dereferencing the chare as follows:
@a1[i].A(); a1[i].B();@

In addition to single method invocations, messages can be broadcasted
to all of the chares in a chare-array by omitting the chare index when
invoking a chare method:

@a1.doIt(parameters);@

Similarly, reductions can be performed on chare-arrays. This
functionality requires that the participating chares in the
chare-array expose a @contribute@ method with the following
definition:

@void contribute(int nBytes, const void *data,@\\
@CkReduction::reducerType type);@

Some of the built-in reduction types supported by the \charmpp\
language are given in \Cref{tab:redtypes}.

\begin{table}[!ht]
\centering
\tabcolsep=0.11cm
\scalebox{0.85}{
\rowcolors{2}{gray!25}{white}
\begin{tabular}{ll}
  \toprule
  Reduction Type & Description \\
  \midrule
nop & no operation performed\\
sum\_int, sum\_float, sum\_double & sum of the given numbers\\
product\_int, product\_float, product\_double & product of the given numbers\\
max\_int, max\_float, max\_double & largest of the given numbers\\
min\_int, min\_float, min\_double & smallest of the given numbers\\
logical\_and & logical AND of the given integers\\
logical\_or & logical OR of the given integers\\
bitvec\_and & bitvector AND of the given numbers\\
bitvec\_or & bitvector OR of the given numbers\\
  \bottomrule
\end{tabular}
}
\caption{Built-in reduction types in \charmpp.}
\label{tab:redtypes}
\end{table}

\paragraph{Structured Dagger}

Structured Dagger (SDAG) is a coordination language built on top of
\charmpp\ that facilitates a clear expression of control flow by
specifying constraints for message-driven
execution~\cite{StructDaggerEURO96}. Consider an example to multiply a
row and column of a matrix:

\begin{figure}[!h]
\centering
\begin{lstlisting}
     // in .ci file
     chare Mult {
       entry void Mult();
       entry void recvRow(Input r);
       entry void recvCol(Input c);
     };
     // in C++ file
     class Mult : public CBase_Mult {
       int   count;
       Input row, col;
     public:
       Mult() {
         count = 2;
       }
       void recvRow(Input r) {
         row = r;
         if (-count == 0) multiply(row, col);
       }
       void recvCol(Input c) {
         col = c;
         if (-count == 0) multiply(row, col);
       }

\end{lstlisting}
\caption{\charmpp\ program for multiplying a row and a column for a matrix.}
\label{fig:charm-mult}
\end{figure}

\begin{figure}[!h]
  \centering
  \begin{lstlisting}
       // in .ci file
       chare Mult {
         entry void Mult();
         entry void recv() {
           when recvRow(Input r)
           when recvCol(Input c)
           serial {
             multiply(r, c);
           }
         entry void recvRow(Input r);
         entry void recvCol(Input c);
       };
       // in C++ file
       class Mult : public CBase_Mult {
         Mult_SDAG_CODE
       public:
         Mult() { recv(); }
  \end{lstlisting}
  \caption{The same \charmpp\ program using SDAG for multiplying a row and a
    column of a matrix.}
  \label{fig:charm-mult-sdag}
  \end{figure}

In this program, the triggers for @multiply@ are constrained using a
member variable @count@. This approach can obfuscate the flow of
control and is potentially error-prone. The basic constructs of SDAG
provide for program-order execution of the entry methods and code
blocks that they define.

SDAG allows the code referenced in \Cref{fig:charm-mult} to be
transformed to that shown in \Cref{fig:charm-mult-sdag} allowing a
clear expression of the constraints that trigger a method. SDAG
introduced keywords @serial@ for atomic blocks, @when@ for conditional
method execution, @overlap@ to process triggers in any order, @forall@
which is equivalent to a ``parallel for'', and finally a @case@
construct to express a disjunction over a set of @when@ clauses.

A complete example of a \charmpp\ program to compute the \textit{n}th
Fibonacci value is shown in \Cref{fig:charm-fib}. To allow further
locality-aware optimizations in the programming model, the \charmpp\
language provides @group@ and @node-group@ constructs. These provide
the facility to create a collection of chares with a single chare on
each PE (in case of groups) or process/logical node (for node groups).

\begin{figure}[!h]
\centering
\begin{minipage}[t]{0.98\textwidth}
\begin{lstlisting}
// fib.ci
mainmodule fib {
  mainchare Main {
    entry Main(CkArgMsg* m);
  };
  chare Fib {
    entry Fib(int n, bool isRoot, CProxy_Fib parent);
    entry void calc(int n) {
      if (n < 3) atomic { respond(seqFib(n)); }
      else {
        atomic {
          CProxy_Fib::ckNew(n - 1, false, thisProxy);
          CProxy_Fib::ckNew(n - 2, false, thisProxy);
        }
        when response(int val)
          when response(int val2)
            atomic { respond(val+val2); }
      }
    };
    entry void response(int val);
  }; };
// fib.cc
struct Main : public CBase_Main {
  Main(CkArgMsg* m) { CProxy_Fib::ckNew(atoi(m->argv[1]), true, CProxy_Fib()); } };

struct Fib : public CBase_Fib {
  Fib_SDAG_CODE
  CProxy_Fib parent; bool isRoot;
  Fib(int n, bool isRoot_, CProxy_Fib parent_)
    : parent(parent_), isRoot(isRoot_) {
    calc(n);
  }
  int seqFib(int n) { return (n < 2) ? n : seqFib(n - 1) + seqFib(n - 2); }
  void respond(int val) {
    if (!isRoot) {
      parent.response(val);
      delete this;
    } else {
      CkPrintf("Fibonacci number is: %d\n", val);
      CkExit();
    }
  } };
\end{lstlisting}
\end{minipage}
\caption{\charmpp\ program to compute the \textit{n}th Fibonacci number.}
\label{fig:charm-fib}
\end{figure}

\subsection{Execution Model}

A basic unit of parallel computation in Charm++ programs is a
\textit{chare}. Chares primarily communicate with each other using
messages. These message object can be user-defined through arbitrary
pack and unpack methods. The entry methods of a chare can be remotely
invoked. They are asynchronous and non-preemptive. The \charmpp\
execution model supports three types of objects:
\begin{itemize}
\item Sequential objects (regular methods).
\item Concurrent objects (chare entry methods).
\item Replicated objects (chare group and node-group objects).
\end{itemize}

In the Charm runtime model~\cite{CharmSys2TPDS94}, chare creation
happens asynchronously. The scheduler picks a message, creates a new
chare if the message is a seed (i.e. a constructor invocation) for a
new Chare, and invokes the method specified by the message. As seen
previously, chares can be grouped into collections. The types of
collections of chares supported in \charmpp\ are: chare-arrays,
chare-groups, and chare-nodegroups.

Chare-arrays are mapped to processors according to a user-defined map
group.  A group is a collection of replicated chares, with exactly one
member element on each processing element. Similarly, a node-group has
one member element on each process or logical node.

Each \charmpp\ program has a main-chare that is invoked after the
\charmpp\ runtime is bootstrapped. This main-chare initializes all the
read variables declared in all of the chares. The main-chare
constructor starts the computation by creating arrays, other chares,
and groups.

To make chare methods serializable and to generate a global object
space, the methods are declared in a separate interface file
``.ci''. A preprocessing step generates proxy classes for each chare
class. These proxy classes act as forwarders at runtime to route the
messages to the appropriate chares.

\begin{table}[!ht]
\centering
\rowcolors{2}{gray!25}{white}
\tabcolsep=0.15cm
\scalebox{0.85}{
\begin{tabular}{lp{16cm}}
  \toprule
  Policy & Description \\
  \midrule
  RandCentLB & Random assignment of objects to processors\\
  GreedyLB & Greedy algorithm to assign heaviest object to the least-loaded processor\\
  GreedyCommLB & Extends greedy algorithm to account for the object communication graph\\
  MetisLB & Uses METIS to partition object-communication graph\\
  TopoCentLB & Extends greedy algorithm to account for processor topology\\
  RefineLB & Minimizes object migration by moving objects away from the most-loaded processors\\
  RefineSwapLB & Same as RefineLB. But when when migration fails, it swaps objects to reduce load on the most-loaded processor\\
  RefineCommLB & Same as RefineLB but accounts for communication\\
  RefineTopoLB & Same as RefineLB but accounts for processor topology\\
  ComboCentLB & Used to combine more than one centralized load balancers\\
  NeighborLB & Neighborhood-aware LB where each processor averages out its load among its neighbors\\
  WSLB & A load balancer for workstation clusters, which can detect load changes and adjust load without interfering interactive usage\\
  \bottomrule
\end{tabular}
}
\caption{Load-balancing policies in the \charmpp\ execution model.}
\label{tab:lbpolicies}
\end{table}

\charmpp\ recognizes two logical entities: a PE (processing element)
and a ``logical node''. The \charmpp\ runtime is divided into several
logical nodes (denoted by processes) running on actual physical
nodes. Each logical node might have several processing elements (PE).
In a \charmpp\ program, a PE is a unit of mapping and scheduling: each
PE has a scheduler with an associated pool of messages. Each chare
resides on one PE at a time, and all PEs within the logical node share
the same memory address space. The PEs continually run a scheduler
that implements several load-balancing policies to manage the load in
the system. These schedulers also poll for messages from the network
and enqueue methods based on the arrival of messages. \charmpp\
supports threaded entry points where each of the methods can be
launched in a separate light-weight thread, synchronizing with its
return value through a future.

Based on the available runtime metrics, \charmpp\ implements several
centralized and distributed load-balancing
schemes. \Cref{tab:lbpolicies} gives a brief description of all the
available load-balancing policies.

Charm++ also supports automatic checkpoint/restart, as well as fault
tolerance based on distributed checkpoints. Experimental GPU support
and shared-memory optimizations have also been
implemented~\cite{KdTree}.

\subsection{Implementation}

The latest stable release, \charmpp\ 6.5.1, supports a variety of
underlying hardware architectures including the BlueGene/L,
BlueGene/P, BlueGene/Q, Cray XT, XE and XK series (including XK6 and
XE6), a single workstation or a network of workstations (including x86
running Linux, Windows, MacOS). Several network backends have been
implemented including UDP, TCP, Infiniband, Myrinet, MPI, uGNI, and
PAMI.

The implementation also includes extensions to the original model
through composition frameworks such as Converse~\cite{InterOpIPPS96},
domain-specific static dataflow languages like Charisma~\cite{orch06},
and libraries such as AMPI~\cite{ampi03} which allow seamless
migration of legacy applications.

The \charmpp\ parallel programming model has been widely adopted, and
has demonstrably proven the advantages of asynchronous message-driven
execution models. Many applications, including the molecular dynamics
application NAMD~\cite{NamdSC02}, an N-body solver
ChaNga~\cite{2009ChaNGaGPU} have shown both productivity and
performance benefits. Besides application, several tools to maximize
programmer productivity such as visualization tools, debugging tools
and simulators (BigSim~\cite{BgSimIPDPS04}) are available for the
\charmpp\ ecosystem.

\section{\parallex}
\label{sec:parallex}

\parallex\ is an execution model for programming large-scale dynamic,
adaptive applications for exascale-class supercomputing
systems~\cite{Parallex}. The main emphasis of \parallex\ is on
avoiding global synchronization whenever possible, under the
contention that it limits scalability now and will limit it further on
future systems. \parallex\ provides a globally-accessible address
space, the Active Global Address Space, allowing regions of memory and
computation to be moved as necessary to increase performance.
The \parallex\ model also includes light-weight, user-level threads
with fast context-switching that allow applications to spawn millions
of threads for hiding latency. Local Control Objects (LCOs), such as
futures, are used for synchronization. Communication uses parcels, a
form of asynchronous active message; when a parcel is received, a
particular method or function is invoked on its target object. Global
identifiers are used to name objects and memory locations in a manner
that allows a common namespace across distinct physical address
spaces. Parcels include continuations, which determine what should be
done with the result of the parcel's action if there is one.
\parallex\ also includes percolations, which allow automatic staging
of data to and from accelerators such as GPUs, as well as support for
fault tolerance as is expected to be necessary for the effective use
of exascale systems. \parallex\ is designed to enable strong scaling
of applications, allowing them to exploit the full capability of the
large-scale parallel systems (including exascale) anticipated by the
end of this decade.

\begin{figure}[!ht]
  \centering
  \noindent\begin{minipage}{.98\textwidth}
  \begin{lstlisting}
  boost::uint64_t fibonacci(boost::uint64_t n);
  HPX_PLAIN_ACTION(fibonacci, fibonacci_action);
  
  boost::uint64_t fibonacci(boost::uint64_t n)
  {
      if (n < 2) return n;
  
      // execute the Fibonacci function locally.
      hpx::naming::id_type const locality_id = hpx::find_here();
  
      using hpx::lcos::future;
      using hpx::async;
  
      fibonacci_action fib;
      future<boost::uint64_t> n1 = async(fib, locality_id, n-1);
      future<boost::uint64_t> n2 = async(fib, locality_id, n-2);
      return n1.get() + n2.get();
  }
  \end{lstlisting}
  \end{minipage}
  \caption{A \hpx\ program to compute the \textit{N}th fibonacci value.}
  \label{fig:hpx-fib}
  \end{figure}

\subsection{Programming Model}

Although there are concrete realizations of the \parallex\ execution
model, such as \hpx, these were largely experimental runtime systems
developed for early performance evaluation of dynamic applications
executed in a message-driven, latency-tolerant fashion. These runtime
systems do not leverage the full potential exposed by the \parallex\
execution model. A novel interface, \xpi\ (\parallex\ Programming
Interface), exposes a set of library calls that interface directly
with the \parallex\ execution model. While sufficiently high-level to
program with, this interface is meant to be a target to high-level
programming models designed to take advantage of the \parallex\
execution model.

In this section, we discuss the \xpi\ programming interface, and the
\hpx\ programming model. We defer the discussion of a language
extension and syntactic constructs to access the active global
address-space to \Cref{sec:discussion}.

\begin{figure}[!ht]
\centering
\begin{lstlisting}
XPI_register_action("fib", fib, 0);
XPI_register_action("set_future", future, 0);

XPI_Err fib(XPI_Addr addr, int n) {
  XPI_Addr f1, f2;
  int nn;
  XPI_Future_new_sync(sizeof(int64_t), &f1);
  XPI_Future_new_sync(sizeof(int64_t), &f2);

  nn = n-1;
  XPI_Parcel p;
  XPI_Parcel_create(&p);
  XPI_Parcel_set_addr(p, addr);
  XPI_Parcel_set_action(p, "fib");
  XPI_Parcel_set_data(p, sizeof(int), &n1);
  XPI_Parcel_set_cont_addr(p, addr);
  XPI_Parcel_set_cont_action(p, "set_future");
  XPI_Parcel_set_cont_data(p, sizeof(XPI_Addr), &f1);
  XPI_Parcel_send(p, XPI_NULL);
  nn = n-2;
  XPI_Parcel_set_data(p, sizeof(int), &n1);
  XPI_Parcel_set_cont_data(p, sizeof(XPI_Addr), &f2);
  XPI_Parcel_free(p);

  uint64_t x, y;
  XPI_Thread_wait_all(f1, f2);
  XPI_Future_get_value_sync(f1, &x);
  XPI_Future_get_value_sync(f1, &y);
  return x+y;
}
\end{lstlisting}
\caption{A \xpi\ program to compute the \textit{N}th fibonacci value.}
\label{fig:xpi-fib}
\end{figure}

The code listing in \Cref{fig:xpi-fib} demonstrates a simple Fibonacci
program written using \xpi. \Cref{fig:hpx-fib} shows the same program
written in \hpx. While the \xpi\ programming model is more verbose as
a result of being implemented for a restricted language C, it also
offers more control over the underlying \parallex\ execution
model. The \xpi\ program shows an example where two future LCOs are
allocated, and filled by continuation parcels using split-phase
transactions. All of the children actions would, presumably, execute
at a single locality if the AGAS address @addr@ were not to move
during the entire execution. By allocating an AGAS array using
user-defined distribution hints, an explicit control over the location
of \parallex\ threads can be obtained. However, this style is
discouraged by the \parallex\ programming paradigm, as the runtime
would detect and migrate threads (and the corresponding AGAS
addresses) when it detects a hot-spot on a particular locality.

As a programming model, \parallex\ allows creation of threads to
express logical parallelism in the program. Threads belonging to the
same locality can communicate and synchronize using Local Control
Objects (LCOs). Arbitrary threads can communicate through the Active
Global Address Space (AGAS). Two threads cannot communicate using
message-passing, however threads can spawn an arbitrary number of
children threads by sending parcels (active messages).

\subsection{Execution Model}

The ``SLOW'' model of performance that highlights four related factors,
each of which acts as a potential source of performance
degradation. \emph{Starvation} occurs when there is insufficient
concurrent work to be performed by computing resources (e.g.,
processor cores), either because the total amount of work is
insufficient or that the distribution of work is uneven with some
resources oversubscribed and others under-utilized. \emph{Latency} is
the distance of an access or service request, often measured in
processor clock cycles. It is the time-distance delay intrinsic to
accessing remote resources and services. \emph{Overhead} is work
required to manage parallel actions and resources on the critical-path
that would be unnecessary in a purely sequential execution. Finally,
contention (\emph{Waiting}) for shared logical or physical resources
can further limit scalability and cause performance
degradation. Memory bank conflicts, limited network bandwidth,
synchronization objects and other resources used by multiple
requesting agents contribute to this form of delay.

Some of the key features representative of the \parallex\ execution
model are:

\begin{itemize}[leftmargin=*]
\item Split-phase transactions using parcels and continuations.
\item Message-driven execution.
\item Distributed shared-memory (not cache coherent) with migration.
\item Local Control Objects for synchronization.
\item Percolation (pre-staging of task data).
\end{itemize}

\paragraph{Lightweight Threads}
\parallex\ represents unit of computations using fine-grained actions
referred to as ``threads''. These have a separate stack, and are
typically implemented as lightweight threads. The \parallex\ execution
model assumes simultaneous execution of hundreds of thousands of such
parallel actions. Threads are ephemeral and can be created and
destroyed at runtime. Threads have an address in the global
address-space. When a thread is blocked, the \parallex\ scheduler
quickly switches to a \textit{ready} thread based on a local
scheduling policy. The existence of several of these threads mitigates
the issues of starvation and latency while maintaining a higher
utilization of the available parallel resources.

\paragraph{Parcels and Message-driven Computation}
Parcels are a form of asynchronous active message that enable
message-driven computation. They are like ``active messages'' as they
allow not just data to move to the work, but also work to move to the
data. When a parcel is received, the target locality spawns a thread
represented by the parcel's action on the target object encoded as the
parcel's payload. A parcel can also include a continuation action
which determines what is to be done with the target action's return
value if there is one. Parcels manage latency by making programs more
asynchronous. A special case of parcels, ``percolation'' establishes
tasks to be performed by an accelerator such as a GPU while
overlapping the communication and staging time on the accelerator with
other operations.

\begin{figure}[!ht]
\centering
\includegraphics[scale=0.6]{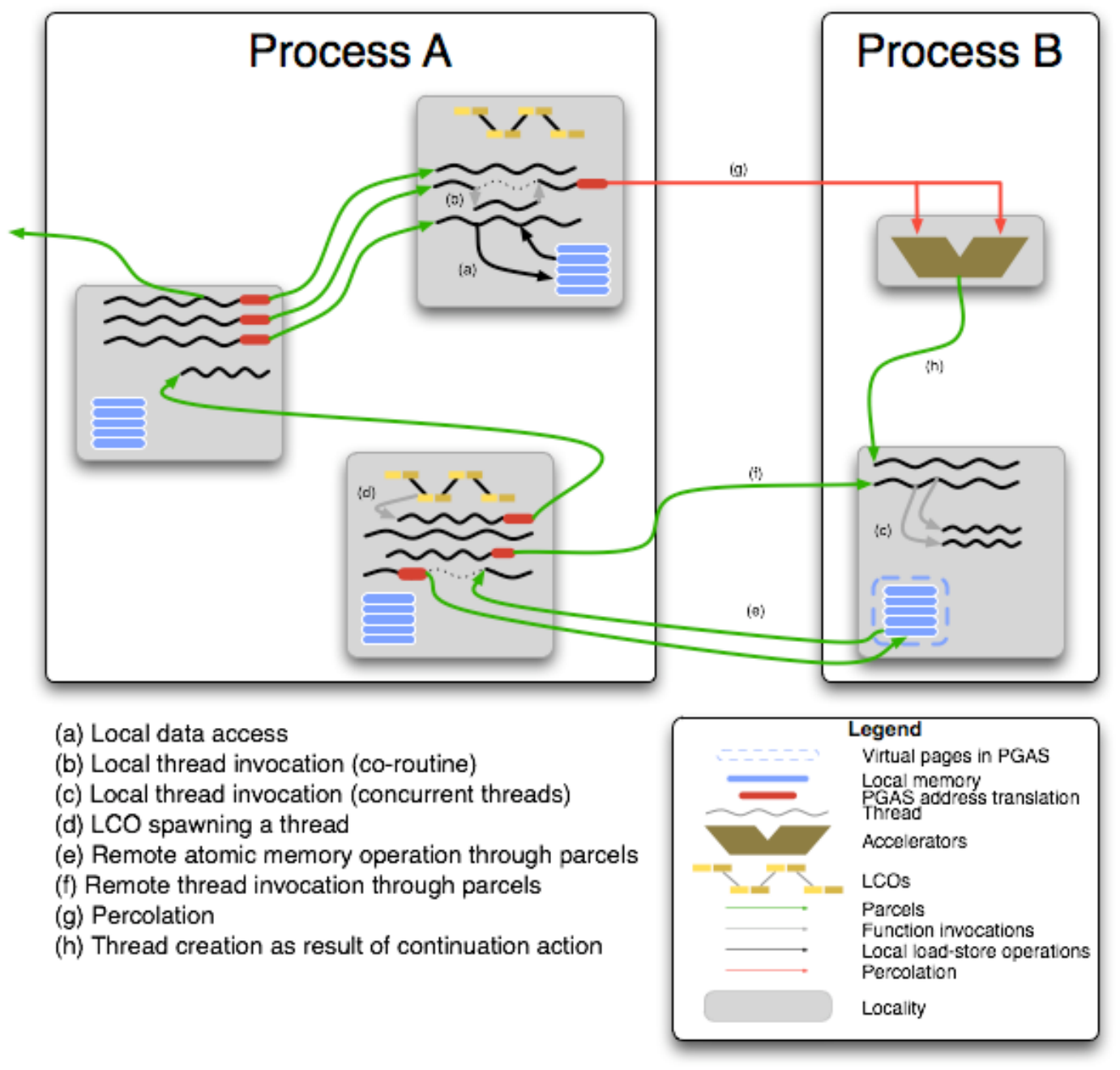}
\caption{Process Interaction in the \parallex\ Model.}
\label{fig:px}
\end{figure}

\paragraph{Active Global Address-Space}
ParalleX defines a globally distributed shared-memory address-space,
which is accessed asynchronously through parcels. It is ``active'' in
the sense that the physical location mapped for a given virtual
address can be changed dynamically by moving data around through
parcels, but keeping the addresses the same. The AGAS offers all of
the benefits of the PGAS models, while allowing addresses to
move. This is done at the expense of a complicated implementation,
possibly, relying heavily on runtime and operating system
support. AGAS defines a distributed environment spanning multiple
localities and need not be contiguous.

\paragraph{Local Control Objects}
Local Control Object (LCOs) are sophisticated synchronization
constructs that encapsulate the control flow of threads. They not only
provide a unifying abstraction to common synchronization primitives
such as semaphores and mutexes, but also allow powerful
synchronization primitives such as futures and dataflow variables.

\paragraph{ParalleX Processes}
The ParalleX execution model organizes its global state in a hierarchy
of context modules referred to as ``ParalleX processes''. Put simply,
processes provide an abstraction of a parallel job by managing the
application threads, the child processes and the contextual mapping
information. It also incorporates capabilities based access rights for
protection and security.

\Cref{fig:px} shows the interactions between the different aspects of
the \parallex\ execution model to enable message-driven computation.

\subsection{Implementation}

A few open-source prototypical implementations of the \parallex\ execution
model exist. Additional highly-tuned implementations of the execution model are
also currently under development. These implementations strive to provide the
\parallex\ vision and philosophy while demonstrating good performance at higher
scales.

\subsubsection{HPX 3}
\begin{figure}[!ht]
\centering
\includegraphics[scale=0.35]{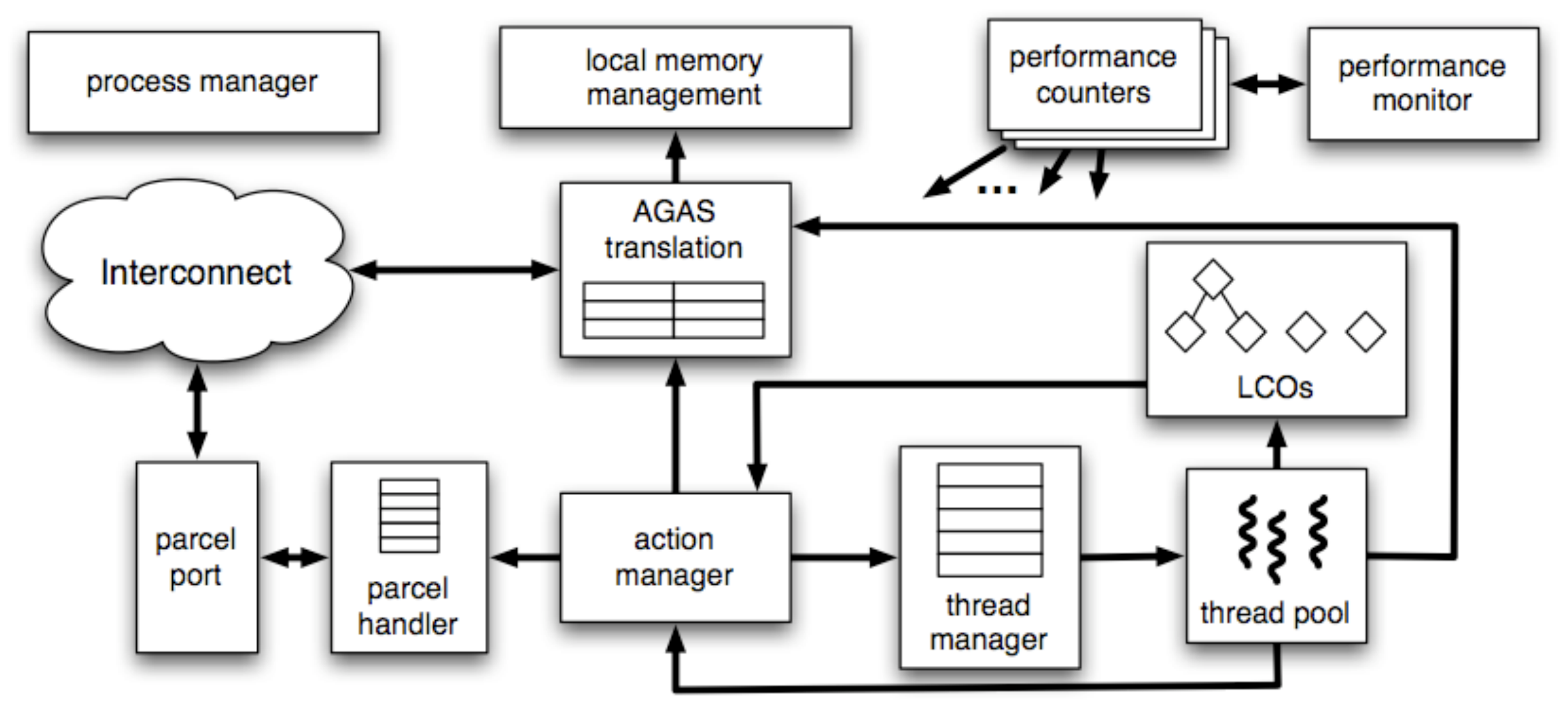}
\caption{The \parallex\ runtime system.}
\label{fig:hpx}
\end{figure}

\Cref{fig:hpx} shows an abstract architecture for an implementation of
the \parallex\ adaptive runtime system. The software architecture of
\hpx\ is very similar to the abstract architecture shown in
\Cref{fig:hpx}. \hpx\ is currently being developed at The STE||AR
Group at Louisiana State University. It is developed as a library on
top of \cpp. \hpx\ does not realize the full potential of
the \parallex\ execution model with several key features currently
unimplemented. Finally, it presently supports limited network backends
with a notable absence of high-performance network interconnection
support.

\subsubsection{HPXi}
HPXi is a prototype implementation in C++11 of a simplified subset of
the full ParalleX model.  The main omission is the Active Global
Address Space feature of ParalleX; objects in HPXi are instead locked
to the address space that created them.  This simplifies addressing
and increases performance, but loses some aspects of the dynamic
adaptivity of the ParalleX model.  HPXi has a global space of objects,
but raw memory is not globally accessible and explicit action is
required to expose an object to remote calls from other address
spaces. HPXi also provides both TBB-style lightweight tasks (without
stacks) in addition to the more capable user-level threads in the
ParalleX model.  For simplicity and ease of implementation, HPXi is
additionally missing the fault-tolerance capabilities that are
integral to the ParalleX model, especially on the planned
less-reliable exascale systems; it is also missing support for dynamic
spawning of new processes and accelerators, and the handling of
responses to parcels must be done manually by the user, unlike the
automatic use of continuations in the full ParalleX model.

As HPXi is a prototype designed to work on current systems, it uses
asynchronous MPI point-to-point operations as its underlying
communication infrastructure.  This aspect enables easy access to
high-performance interconnection networks, although it may have higher
overheads than a lower-level interface such as GASNet or writing
drivers for particular types of network hardware.  Although MPI is
used as infrastructure, HPXi wraps that with an active message
("parcel") interface, not send/receive or collective operations.  HPXi
also uses existing mechanisms to create system-level threads, building
lighter-weight task and user-level thread abstractions on top of those
using Boost.Context and Boost.Coroutine.

HPXi includes several features to ease porting of legacy MPI and
MPI-style applications.  For example, it includes MPI-style send and
receive operations, as well as various forms of collective
communication.  Unlike standard MPI, it allows multiple objects in the
same address space (and thus underlying MPI rank) to act as different
emulated MPI processes. These operations interoperate with user-level
threads to enable non-blocking, asynchronous use of blocking MPI-style
code.  HPXi also includes multiple forms of parallel for loops (both
for independent iterations and reduction operations) to ease porting
of simple OpenMP and TBB programs.  Both TBB-style spawning of a
precomputed number of tasks and Cilk-style recursive decomposition of
a sequence into dynamic tasks as necessary are supported.  These
features, plus their relatively low performance overheads, allow
incremental porting of legacy applications, removing the need to
rewrite them completely into a fully asynchronous, object-based
programming model all at once; however, this porting process is
necessary longer-term to achieve the best scalability of the
applications on extreme-scale systems.

\subsubsection{HPX 5}

HPX 5 is a high-performance, production-capable library in C,
currently being developed at CREST at Indiana University. It is
planned to have close integration with the operating system (through
the RIOS interfaces), an optimized parcel communication layer built on
top of high-performance one-sided interfaces and efficient threading
and synchronization implementations. It supports a byte-addressable,
active global address space (AGAS) for storing distributed data
objects.

\section{Active Pebbles and \ampp}
\label{sec:pebbles}

Active Pebbles (AP) is a novel parallel programming model suited for
data-driven problems that are fine-grained, irregular and
non-local. The model provides a marked distinction between the natural
expression of an algorithm, and its underlying efficient execution
mechanisms. The corresponding execution model is implemented using a
new user-level library, \ampp, for programming with high-level active
messages based on generic programming principles. The efficacy of the
model has been demonstrated by succinctly and directly expressing
graph algorithms and other irregular application kernels. The
experimental results exhibit performance comparable to MPI-based
implementations that are significantly more complicated.

\subsection{Programming Model: Active Pebbles}

The key elements in the Active Pebbles model are: \emph{pebbles},
which are light-weight active messages (AM) managed and scheduled by
the runtime; \emph{handlers}, which are functions that are executed on
\emph{targets} in response to pebbles (or ensembles of pebbles); and
\emph{distribution objects} that dictate the distribution of data
objects.

The four main techniques employed by the model are:
\begin{enumerate}[leftmargin=*]
\item Global address-space using fine-grained pebble addressing.
\item Optimized communication using active hypercube routing.
\item Communication reduction using message coalescing and message reduction.
\item Termination detection.
\end{enumerate}

\begin{figure}[!ht]
\centering
\includegraphics[scale=0.35]{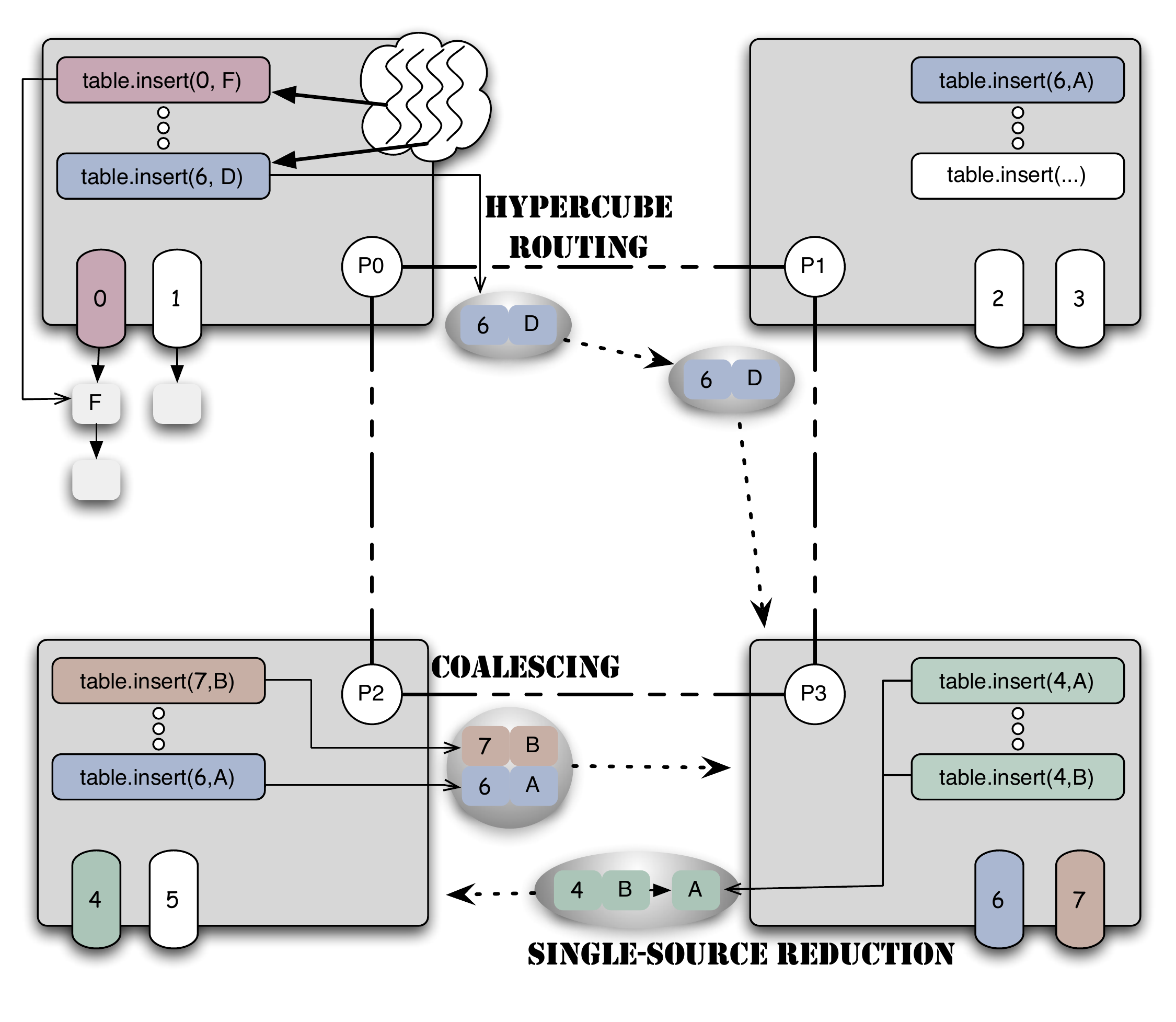}
\caption{The Active Pebbles Programming Model.}
\label{fig:ap}
\end{figure}

Active messages help to migrate computation to data, and are mostly
used in low-level communication layers. Programming with active
messages can be cumbersome and error-prone owing to the unnatural
restructuring of the algorithm in a continuation-passing style. Active
Pebbles raises the abstraction of programming with active messages as
the handlers can be more coarser grained, and the runtime handles
optimizations such as coalescing, reduction and termination detection.

The programmer defines pebbles -- message types that encapsulate the
data and its handler. Computation is performed by addressing pebbles
to targets mapped to different nodes. Active pebbles generalizes
one-sided operations to user-defined pebble handlers, thereby removing
the need for expensive pairwise synchronization. Pebble coalescing
ensures that the pebbles addressed to the same node are grouped
together to increase message sizes. Intermediary nodes can potentially
perform reductions on coalesced pebbles to eliminate duplicates or
combine idempotent operations. These optimizations are admitted due to
active routing over a hypercube overlay topology where multiple nodes
are involved in transferring a pebble from its source to its
destination. These optimizations are central to the model and form the
basis of the Active Pebbles programming model.

\begin{figure}[!htb]
\centering
\begin{lstlisting}
struct fib::fib_handler {
  fib_handler(fib& self) : self(&self) {
  void operator()(const amplusplus::transport::rank_type src, const fib_data& data) const {
    if(data.first == 1 || data.first == 2) {
      self->response_message.send(std::make_pair(1, data.second), src);
    } else {
      self->divide(data, src);
    }
  }
  fib* self;
};

struct fib::response_handler {
  response_handler(fib& self) : self(&self) {}
  void operator()(const amplusplus::transport::rank_type src, const fib_data& data) const {
    self->merge(data);
  }
  fib* self;
};
\end{lstlisting}
\caption{Handlers for an \ampp\ program to compute the Fibonacci number.}
\label{fig:ampp-fib}
\end{figure}

\subsection{Execution Model: \ampp\ and Implementation}

\ampp\ is an active message framework based on generic programming
techniques. As such, the \ampp\ library is an implementation for an
execution model based on the AP programming model. It raises the
abstraction for the low-level active messages often used by
communication libraries. In particular, the active message handlers
are not forced to run in an interrupt context, and can send arbitrary
messages to other nodes. As \ampp\ is implemented as a C++ library, it
allows statically-typed messages and generic handlers for messages. In
addition to supporting all of the optimizations mandated by the AP
programming model, the \ampp\ library enables compiler optimizations
such as message vectorization and inlining whenever admissible. The
performance of \ampp\ has been shown to be competitive to lower-level
active message libraries such as GASNet and others. 

\section{Comparison}
\label{sec:comparison}

\Cref{tab:comp-pm} compares the Cilk, \charmpp\, \parallex\ and the Active
Pebbles programming models. \emph{Task Parallelism} is the capability of
expressing computation as task leveraging fine-grained concurrency within the
programming model. The means of task creation and synchronization for the three
programming models are given. Further distinction is made on the ability to
attach explicit continuation tasks to computation tasks; expressing static
dataflow parallelism through dataflow graphs; determinism and repeatability of
computation. The memory model backing the programming model, and the means to
specify the distribution of data amongst processors is additionally an
important classification criteria, especially for enabling data locality
optimizations. Support for vectorizations, reductions and other aggregate
operations permitted on objects in the shared address space is key to
maximizing parallel work. \emph{Reuse and Modularity} refers to the ability of
the programming model to provide standalone libraries; its amenability for
separate-compilation, and composability with other, potentially non-parallel,
modules.\\

%

\begin{table*}[!ht]
\centering
\noindent\begin{minipage}{1\textwidth}
\renewcommand{\arraystretch}{1.2}
\begin{tabular}{r| >{\centering\arraybackslash}m{2.5cm} | >{\centering\arraybackslash}m{2.5cm} | >{\centering\arraybackslash}m{2.5cm} | >{\centering\arraybackslash}m{2.5cm}}
  \hline
  & \heading{Cilk} & \heading{\charmpp} & \heading{\parallex} & \heading{Active Pebbles} \\
  \hline
  Task Parallelism              & \Yes\; & \Yes\; & \Yes\; & \Yes\; \\
  Task Creation                 & {\lstinline!cilk_spawn!} and {\lstinline!cilk_for!} & {\lstinline!ckNew!} (chare creation) & {\lstinline!hpx_call!} (thread creation) & \emph{handlers} \\
  Task Synchronization          & {\lstinline!cilk_sync!} and implicit & SDAG & LCO & - \\
  Task Execution                & Asynchronous & Asynchronous & Asynchronous & Asynchronous \\
  Explicit continuations        & \No\parnote{Obsolete} & \No\; & \Yes\; & \Maybe\; \\
  Dataflow parallelism          & \No\; & \Yes\; & \Yes\; & \No\; \\
  Determinism and repeatability & \Yes\; & \Maybe\parnote{Write-once objects} & \Yes\parnote{LCOs} & \No\; \\
  Data Parallelism              & \Yes\parnote{Array notation, Elemental functions} & \Yes\parnote{Chare collections} & \Yes\parnote{AGAS array distributions} & \Yes\parnote{Distribution Objects} \\
  Memory                        & Shared-Memory & Shared-Memory; Global Object Space & Shared-Memory; Global Address Space & Shared-Memory; Message-Passing \\
  Explicit Data Distribution    & \No\; & \Yes\; & \Yes\; & \Yes\; \\
  Vectorized Operations         & \Yes\; & \No\; & \No\; & - \\
  Reductions                    & \Yes\parnote{Reducer objects} & \Yes\parnote{Reduction on chare collections} & \Yes\parnote{Using LCOs} & \Yes\parnote{Pebble Reduction} \\
  Sequential Semantics          & \Yes\; & \No\; & \No\; & \No\; \\
  Reuse and modularity          & \Yes\; & \Maybe\; & \Maybe\; & \Maybe\; \\
  \hline
\end{tabular}
\parnotes
\parnotereset
\end{minipage}
\caption{Comparison of the programming models.}
\label{tab:comp-pm}
\end{table*}

\Cref{tab:comp-em} compares the execution models of Cilk, \charmpp\, \parallex\
and \ampp. The criteria used for comparison is primarily based on the execution
semantics as defined by the programming model, and the features offered by its
specific implementation under consideration. We consider the unit of
computations, their representation and implementation details such as
fine-grained concurrency, hierarchical parallelism, message-driven computation
and multi-node execution. Furthermore, synchronization between tasks and their
scheduling plays an important role in minimizing parallel overheads at scale.
Dynamic behaviors of runtime systems such as adaptive parallelism, distributed
load balancing, automatic redistribution of data are key for sustained
throughput and performance of irregular applications at scale. Other features
such as fault tolerant, parallel I/O and support for accelerators are also
considered.\\


\begin{table*}[!h]
  \centering
  \noindent\begin{minipage}{1\textwidth}
  \renewcommand{\arraystretch}{1.2}
  \begin{tabular}{r| >{\centering\arraybackslash}m{2.5cm} | >{\centering\arraybackslash}m{2.5cm} | >{\centering\arraybackslash}m{2.5cm} | >{\centering\arraybackslash}m{2.5cm}}
    \hline
    & \heading{Cilk} & \heading{\charmpp} & \heading{\parallex} & \heading{\ampp}\\
    \hline
    Unit of computations          & Tasks & Chares & Actions & Active Messages \\
    Task representation           & Parallel function call & Entry Methods &
    Threads & Handlers \\
    Fine-grained Threading        & \Maybe\; & \Yes\; & \Yes\; & \Yes\; \\
    Message-driven computation    & \No\; & \Yes\parnote{Chare invocations} & \Yes\parnote{Parcel sends} & \Yes\; \\
    Multi-node execution          & \Maybe\; & \Yes\; & \Yes\; & \Yes\; \\
    Hierarchical execution        & \Maybe\parnote{Interoperability
      with a distributed model} & \Yes\; & \Yes\; & \Yes\; \\
    Synchronization between tasks & Locks & SDAG & LCO & - \\
    Task Scheduling               & Work-stealing & Various & Various & Work-sharing \\
    Topology-aware mapping        & \No\; & \Yes\; & \Yes\; & \No\; \\
    Adaptive Parallelism          & \Maybe\; & \Yes\; & \Yes\; & \Yes\; \\
    Automatic Redistribution      & \No\; & \Yes\; & \Yes\; & \Maybe\; \\
    Distributed Shared-Memory     & \Maybe\; & \No\; & \Yes\; & \Yes\; \\
    Global address space          & \No\; & \Yes\; & \Yes\; & \Maybe\; \\
    Quiescence detection          & \No\; & \Yes\; & \Yes\; & \Yes\; \\
    Fault Tolerance               & \Maybe\; & \Yes\parnote{Checkpointing} & \Yes\; & \No\; \\
    Parallel I/O                  & \No\; & \No\; & \Yes\; & \No\; \\
    Accelerator support           & \No\; & \Yes\parnote{GPU integration} & \Yes\parnote{Percolation} & \Maybe\; \\
    \hline
  \end{tabular}
  \parnotes
  \end{minipage}
  \caption{Comparison of the execution models.}
  \label{tab:comp-em}
  \end{table*}

\section{Discussion}
\label{sec:discussion}

To address impending challenges at exascale and beyond, execution models would
need to be asynchronous, dynamic, resilient, adaptive and latency-tolerant. On
top of this, they would need to expose unifying abstractions for data and task
parallelism to programmers while allowing them to strike a balance between
productivity, performance, safety and correctness.

Latency-tolerant optimizations such as split-phase transactions allow the
decoupling of synchronous execution through a continuation-passing style
transformation of the program. Future execution models, including the ones that
were compared in this paper, would need to expose logical parallelism at
varying granularity and allow an intelligent runtime to schedule and execute
the computation adaptively and autonomously.

Performance prediction of applications in distributed dynamic execution models
is an emerging research problem. The overall performance gain of an application
largely depends on the characteristics of the algorithm itself. But limited
improvement in performance is observed already by a straightforward translation
of the BSP-style communication primitives to their equivalent in dynamic
runtime systems such as \charmpp or HPX. This is due entirely to the
finer-grained concurrency allowed by the execution model. Better performance
can be achieved by leveraging more features admitted by dynamic runtime
systems, and at times, through a complete rewrite of the application's
algorithm. At higher scales, performance prediction through modeling and
simulation would be key factor in co-design and implementation of scalable
algorithms.

Our comparison identifies a set of common features present in the HPC runtime
systems under evaluation. These emerging class of runtime systems are centered
around the asynchronous, massively multi-threaded model of computation in which
lightweight threads operate on data residing in globally shared memory. This
model, now commonly referred to as \emph{asynchronous many-tasking}, is
characterized by several independently executing tasks, often in the order of
millions or billions, that are automatically scheduled based on some complex
resource criteria. These tasks are ephemeral and event-driven in nature
triggered by \emph{active messages} exchanged in the system. As explicit
communication tends to be error-prone given the massive amount of exposed
concurrency, AMT models are often coupled with a global address space that
allows sharing data between tasks. Furthermore, as AMT models encompass and
extend the SPMD paradigm relinquishing a substantial amount of execution
control to the runtime system, they are well-suited for exascale-class
applications.


%% file: ms.bbl
\begin{thebibliography}{10}

\bibitem{cuda}
{\em {NVIDIA} {CUDA} Programming Guide}, 2007.

\bibitem{tbb-reference-manual}
{Threading Building Blocks Reference Manual}, 2011.

\bibitem{MSA-APGAS09}
A.~Becker, P.~Miller, and L.~V. Kale.
\newblock {P}{G}{A}{S} in the message-driven execution model.
\newblock In {\em 1st Workshop on Asynchrony in the {P}{G}{A}{S} Programming
  Model {A}{P}{G}{A}{S}}, June 2009.

\bibitem{Blumofe:1996:DDS:645606.661333}
R.~D. Blumofe, M.~Frigo, C.~F. Joerg, C.~E. Leiserson, and K.~H. Randall.
\newblock Dag-consistent distributed shared memory.
\newblock In {\em Proceedings of the 10th International Parallel Processing
  Symposium}, IPPS '96, pages 132--141, Washington, DC, USA, 1996. IEEE
  Computer Society.

\bibitem{cilk-1996}
R.~D. Blumofe, C.~F. Joerg, B.~C. Kuszmaul, C.~E. Leiserson, K.~H. Randall, and
  Y.~Zhou.
\newblock Cilk: An efficient multithreaded runtime system.
\newblock {\em Journal of Parallel and Distributed Computing}, 37(1):55 -- 69,
  1996.

\bibitem{Blumofe:1997:ARP:1268680.1268690}
R.~D. Blumofe and P.~A. Lisiecki.
\newblock Adaptive and reliable parallel computing on networks of workstations.
\newblock In {\em Proceedings of the annual conference on USENIX Annual
  Technical Conference}, ATEC '97, pages 10--10, Berkeley, CA, USA, 1997.
  USENIX Association.

\bibitem{Charles:2005:XOA:1094811.1094852}
P.~Charles, C.~Grothoff, V.~Saraswat, C.~Donawa, A.~Kielstra, K.~Ebcioglu,
  C.~von Praun, and V.~Sarkar.
\newblock X10: an object-oriented approach to non-uniform cluster computing.
\newblock In {\em Proceedings of the 20th annual ACM SIGPLAN conference on
  Object-oriented programming, systems, languages, and applications}, OOPSLA
  '05, pages 519--538, New York, NY, USA, 2005. ACM.

\bibitem{MapReduce}
J.~Dean and S.~Ghemawat.
\newblock {MapReduce}: {S}implified data processing on large clusters.
\newblock {\em Commun. ACM}, 51:107--113, Jan. 2008.

\bibitem{desouza04:_msa}
J.~DeSouza and L.~V. Kal\'e.
\newblock {MSA}: Multiphase specifically shared arrays.
\newblock In {\em Proceedings of the 17th International Workshop on Languages
  and Compilers for Parallel Computing}, West Lafayette, Indiana, USA,
  September 2004.

\bibitem{Frigo:1998:ICM:277650.277725}
M.~Frigo, C.~E. Leiserson, and K.~H. Randall.
\newblock The implementation of the cilk-5 multithreaded language.
\newblock In {\em Proceedings of the ACM SIGPLAN 1998 conference on Programming
  language design and implementation}, PLDI '98, pages 212--223, New York, NY,
  USA, 1998. ACM.

\bibitem{Parallex}
G.~R. Gao, T.~Sterling, R.~Stevens, M.~Hereld, and W.~Zhu.
\newblock Parallex: A study of a new parallel computation model.
\newblock {\em Parallel and Distributed Processing Symposium, International},
  0:294, 2007.

\bibitem{orch06}
C.~Huang and L.~V. Kale.
\newblock Charisma: Orchestrating migratable parallel objects.
\newblock In {\em Proceedings of IEEE International Symposium on High
  Performance Distributed Computing (HPDC)}, July 2007.

\bibitem{ampi03}
C.~Huang, O.~Lawlor, and L.~V. Kal\'{e}.
\newblock Adaptive {MPI}.
\newblock In {\em Proceedings of the 16th International Workshop on Languages
  and Compilers for Parallel Computing (LCPC 2003), LNCS 2958}, pages 306--322,
  College Station, Texas, October 2003.

\bibitem{KdTree}
P.~Jetley and L.~V. Kale.
\newblock Optimizations for message driven applications on multicore
  architectures.
\newblock In {\em 18th annual IEEE International Conference on High Performance
  Computing ({HiPC} 2011)}, December 2011.

\bibitem{2009ChaNGaGPU}
P.~Jetley, L.~Wesolowski, F.~Gioachin, L.~V. Kal{\'e}, and T.~R. Quinn.
\newblock Scaling hierarchical n-body simulations on gpu clusters.
\newblock In {\em Proceedings of the 2010 ACM/IEEE International Conference for
  High Performance Computing, Networking, Storage and Analysis}, SC '10,
  Washington, DC, USA, 2010. IEEE Computer Society.

\bibitem{CharmppOOPSLA93}
L.~{Kal\'{e}} and S.~Krishnan.
\newblock {CHARM++: A Portable Concurrent Object Oriented System Based on C++}.
\newblock In A.~Paepcke, editor, {\em {Proceedings of OOPSLA'93}}, pages
  91--108. {ACM Press}, September 1993.

\bibitem{StructDaggerEURO96}
L.~V. Kale and M.~Bhandarkar.
\newblock {Structured Dagger: A Coordination Language for Message-Driven
  Programming}.
\newblock In {\em Proceedings of Second International Euro-Par Conference},
  volume 1123-1124 of {\em Lecture Notes in Computer Science}, pages 646--653,
  September 1996.

\bibitem{InterOpIPPS96}
L.~V. Kale, M.~Bhandarkar, N.~Jagathesan, S.~Krishnan, and J.~Yelon.
\newblock {Converse: An Interoperable Framework for Parallel Programming}.
\newblock In {\em Proceedings of the 10th International Parallel Processing
  Symposium}, pages 212--217, April 1996.

\bibitem{CharmSys1TPDS94}
L.~V. Kal{\'e}, B.~Ramkumar, A.~B. Sinha, and A.~Gursoy.
\newblock {The CHARM Parallel Programming Language and System: Part I --
  Description of Language Features}.
\newblock {\em Parallel Programming Laboratory Technical Report \#95-02}, 1994.

\bibitem{CharmSys2TPDS94}
L.~V. Kal{\'e}, B.~Ramkumar, A.~B. Sinha, and V.~A. Saletore.
\newblock {The CHARM Parallel Programming Language and System: Part II -- The
  Runtime system}.
\newblock {\em Parallel Programming Laboratory Technical Report \#95-03}, 1994.

\bibitem{Keleher94treadmarks:distributed}
P.~Keleher, A.~L. Cox, S.~Dwarkadas, and W.~Zwaenepoel.
\newblock Treadmarks: Distributed shared memory on standard workstations and
  operating systems.
\newblock In {\em IN PROCEEDINGS OF THE 1994 WINTER USENIX CONFERENCE}, pages
  115--131, 1994.

\bibitem{opencl-spec}
{Khronos OpenCL Working Group}.
\newblock {\em The OpenCL Specification, version 1.1}, September 2010.

\bibitem{Exascale:DARPA}
P.~Kogge, K.~Bergman, S.~Borkar, D.~Campbell, W.~Carlson, W.~Dally, M.~Denneau,
  P.~Franzon, W.~Harrod, K.~Hill, J.~Hiller, S.~Karp, S.~Keckler, D.~Klein,
  R.~Lucas, M.~Richards, A.~Scarpelli, S.~Scott, A.~Snavely, T.~Sterling,
  R.~Stanley, and K.~Yelick.
\newblock {E}xa{S}cale computing study: Technology challenges in achieving
  exascale systems, Sept.~28, 2008.

\bibitem{Top500}
H.~Meuer, E.~Strohmaier, J.~Dongarra, and H.~Simon.
\newblock Top500 report for {J}une 2013, June~13, 2013.

\bibitem{MPI-2.2}
{MPI Forum}.
\newblock {\textsf{MPI}: A Message-Passing Interface Standard. v2.2}, Sept.
  2009.

\bibitem{MPI-3.0}
{MPI Forum}.
\newblock {\textsf{MPI}: A Message-Passing Interface Standard. v3.0}, Sept.
  2012.

\bibitem{OpenMPSpec}
{OpenMP Architecture Review Board}.
\newblock {OpenMP} application program interface.
\newblock Specification, 2011.

\bibitem{Peng00silkroad:a}
L.~Peng, W.~F. Wong, M.~D. Feng, C.~K. Yuen, L.~Peng, W.~F. Wong, M.~D. Feng,
  and C.~K. Yuen.
\newblock Silkroad: A multithreaded runtime system with software distributed
  shared memory for smp clusters.
\newblock In {\em In IEEE International Conference on Cluster Computing
  (Cluster2000}, pages 243--249, 2000.

\bibitem{NamdSC02}
J.~C. Phillips, G.~Zheng, S.~Kumar, and L.~V. Kal{\'e}.
\newblock {NAMD}: Biomolecular simulation on thousands of processors.
\newblock In {\em Proceedings of the 2002 ACM/IEEE conference on
  Supercomputing}, pages {1--18}, Baltimore, MD, September 2002.

\bibitem{tbb}
J.~Reinders.
\newblock {\em {Intel Threading Building Blocks: {O}utfitting C++ for
  Multi-core Processor Parallelism}}.
\newblock O'Reilly Media, July 2007.

\bibitem{opencl}
J.~E. Stone, D.~Gohara, and G.~Shi.
\newblock {OpenCL}: A parallel programming standard for heterogeneous computing
  systems.
\newblock {\em Computing in Science \& Engineering}, 12(3):66--73, May 2010.

\bibitem{upc}
{UPC Consortium}.
\newblock {UPC Language Spec., v1.2}.
\newblock Technical report, {Lawrence Berkeley National Laboratory}, 2005.
\newblock LBNL-59208.

\bibitem{BgSimIPDPS04}
G.~Zheng, G.~Kakulapati, and L.~V. Kal{\'e}.
\newblock Bigsim: A parallel simulator for performance prediction of extremely
  large parallel machines.
\newblock In {\em 18th International Parallel and Distributed Processing
  Symposium (IPDPS)}, page~78, Santa Fe, New Mexico, April 2004.

\end{thebibliography}
